# Spatio-temporal influence of solar activity on global air temperature

Samuel T. OGUNJO[a,*], A. Babatunde RABIU[b]

[a]*Department of Physics, Federal University of Technology, Akure, PMB 704 Akure, Ondo State, Nigeria*

[b]*Centre for Atmospheric Research, National Agency for Space Research and Development, Anyigba, Kogi State.*

---

**Abstract**

Previous studies on the impact and influence of solar activity on terrestrial weather has yielded contradictory results in literatures. Present study presents, on a global scale, the correlation between surface air temperature and two solar activity indices (Sunspot number, 'Rz', and solar radio flux at 10.7, 'F10.7' ) at different time scales during solar cycle 23. Global air temperature has higher correlation values of ±0.8 with F10.7 compared to Rz (±0.3). Our results showed hemispheric delineation of the correlation between air temperature and solar activity with negative correlation in the southern hemisphere and positive correlation in the northern hemisphere. At the onset of the solar cycle, this hemispheric delineation pattern was prevalent, however, an inverse hemispheric delineation was observed at the recession of the solar cycle.

*Corresponding author

*Email address:* stogunjo@futa.edu.ng (Samuel T. OGUNJO)

## 1. INTRODUCTION

The Sun influences weather on Earth through changes in solar irradiance, variability in solar ultraviolet, and effect of galactic cosmic rays (Mufti and Shah, 2011). The direct and indirect impact of solar activity on Earth's climate is important, hence, the need for continuous monitoring of solar activity. It is pertinent to estimate the contribution of solar activity to global warming and climate change. Over the years, scientists have developed indices to measure and quantify solar activity. One of the most common indices for solar activity is the sunspot number, $R_z$. Sunspots are regions of reduced temperature on the Sun photosphere. Sunspot number $R_z$, was one of the earliest proxies for solar activity with data as far back as the 16th century. The solar 10.7 cm radio flux ($F10.7$ ) is defined as the measurement within one hour, of all emissions on the solar disc at a wavelength of 10.7 cm (Tapping, 2013). The $F10.7$, has been found to be a better representation of solar activity than $R_z$ (Mielich and Bremer, 2013; Okoh and Okoro, 2020). A similar, E10.7 index based on the extreme ultraviolet radiation 10.7 cm radio flux has also been proposed (Tobiska et al., 2000). The flare index is a short-lived solar activity to quantify the daily flare activity over a 24-hour period (Kleczek, 1952). The total sunspot area, obtained by measuring the area of each sunspot group, has been proposed as a measure of solar activity (Sarychev and Roshchina, 2006). These indices provide a means to quantify the effect of solar activity on atmospheric parameters.

The impact of solar activity on atmospheric variables has been investigated.

Values of coefficient of determination $r^2$ between $R_z$ and rainfall were found to be 0.89 in Rome using long term data (Thomas, 1993). Correlation coefficient between $R_z$ and rainfall in India has been reported to be in the range -0.14 to +0.29 (Ananthakrishnan and Parthasarathy, 1984), 0.199 (Hiremath and Mandi, 2004), ±0.2 (Chattopadhyay and Chattopadhyay, 2011), −0.41 − 0.55

(Hiremath, 2006), -0.76 to -0.86 (Selvaraj et al., 2009), ≤ 0.35 (Chakraborty and Bondyopadhyay, 1986), and 0.145 (Jain and Tripathy, 1997). The differences in correlation coefficient values in India could be attributed to different time periods, different number of locations, and method of computation considered. In other regions of the world, correlation values between $R_z$ and annual rainfall has been reported as -0.1 for Southern Brazil (Echer et al., 2008), 0.48 − 0.99 for several African countries, (Fleer, 1982), 0.40 in Portugal (Lucio, 2005), -0.10 in Santa Maria (Rampelotto et al., 2012), and 0.24 in Italy (Mazzarella and Palumbo, 1992). Sunspot activities have also been reported to have correlation with lake volumes/water levels, river flows (Mauas et al., 2011), . Furthermore, the impact of $R_z$ has been reported on large scale teleconnection patterns such as El Nino-southern oscillation (ENSO) (Zaffar et al., 2019), North Atlantic Oscillations (Hern´andez et al., 2020; Kuroda et al., 2022), and Pacific Decadal Oscillation (PDO) (Ormaza-Gonz´alez and Espinoza-Celi, 2018).

One of the greatest interest in Sun-Earth relationship, is the influence of solar activity on tropospheric temperature. The solar contribution to tropospheric temperature has been estimated to be 7% (Solomon et al., 2007), 30% (Solanki and Krivova, 2003), 41% (De Jager et al., 2010), and 60% (Scafetta, 2010). The correlation between mean global air temperature and sunspot number has been estimated at 0.27 (Valev, 2006). Correlation values reported between $R_z$ and air temperature at various location and time periods include +0.57 (Sch¨onwiese, 1978), +0.5 (Blanco and Catalano, 1975), -0.42 Rabiu et al. (2005), -0.26 Rabiu et al. (2005), and +0.66 (Echer et al., 2009). Another approach to estimating the influence of solar activity on air temperature is using the length of the solar cycle. Solheim et al. (2012) observed significant negative trend between Norwegian air temperature and length of previous solar cycle but not with the current solar cycle. The study did not give information about the variation of correlation values for different regions of the world. A high correlation was observed between solar cycle length and air temperature in the northern hemisphere (Friis-Christensen and Lassen, 1991). This correlation between solar cycle length and air temperature have also been confirmed at Northern Ireland (Butler and Johnston, 1996), Svalbard at 12-year lag (Solheim et al., 2011), and Qinghai-Xizang railway at 5-year lag (Li et al., 2004). Correlation between $R_z$ and winter temperature has been estimated as −0.3 in Canada (Laing and Binyamin, 2013), +0.42 in Holland (De Jager, 1981), and −0.91 to −0.63 in Bulgaria (Georgieva et al., 2005).

Weather across the world is connected. Previous studies on the relationship between $R_z$ and air temperature have largely focused on aggregated data and specific locations. However, it is imperative to study the interaction of solar on global weather to determine large scale patterns and trends. This makes it difficult to make inference on the global impact of solar activity. The aim of this study is to characterize the relationship between solar and geomagnetic activities and global air temperature at long term, seasonal and annual time scales across the world. This will give insight into the contribution of solar and geomagnetic contributions to climate activities across different regions of the world.

## 2. METHODOLOGY

For this study, $R_z$ and $F10.7$ were used as proxies for solar activity during solar cycle 23 (1997 - 2008). The daily NCEP-NCAR Reanalysis (http://www.psl.noaa.gov/data/gridded/data.ncep.reanalysis.html) air temperature data at 2 m was used. Daily $R_z$ and $F10.7$ values were obtained from the OMNI database (https://omniweb.gsfc.nasa.gov/form/dx1.html). The seasonal consideration were based on the Lloyd's season where J-season includes May, June, July, and August; D-season months are November, December, January, and February; while the E-Season months are March, April, September, and October.

The Spearman correlation ($\rho$) was used in this study. It is defined as

$$\rho = 1 - \frac{6 \sum d_i^2}{n(n^2 - 1)} \tag{1}$$

The values of $\rho$ are in the range ±1. Negative values denotes negative correlations between the two variables which implies that an increase in one variable corresponds to a decrease in the other variable. The significance of the correlation was computed using the two tailed p-value. In this study, all results were considered at 95% confidence interval.

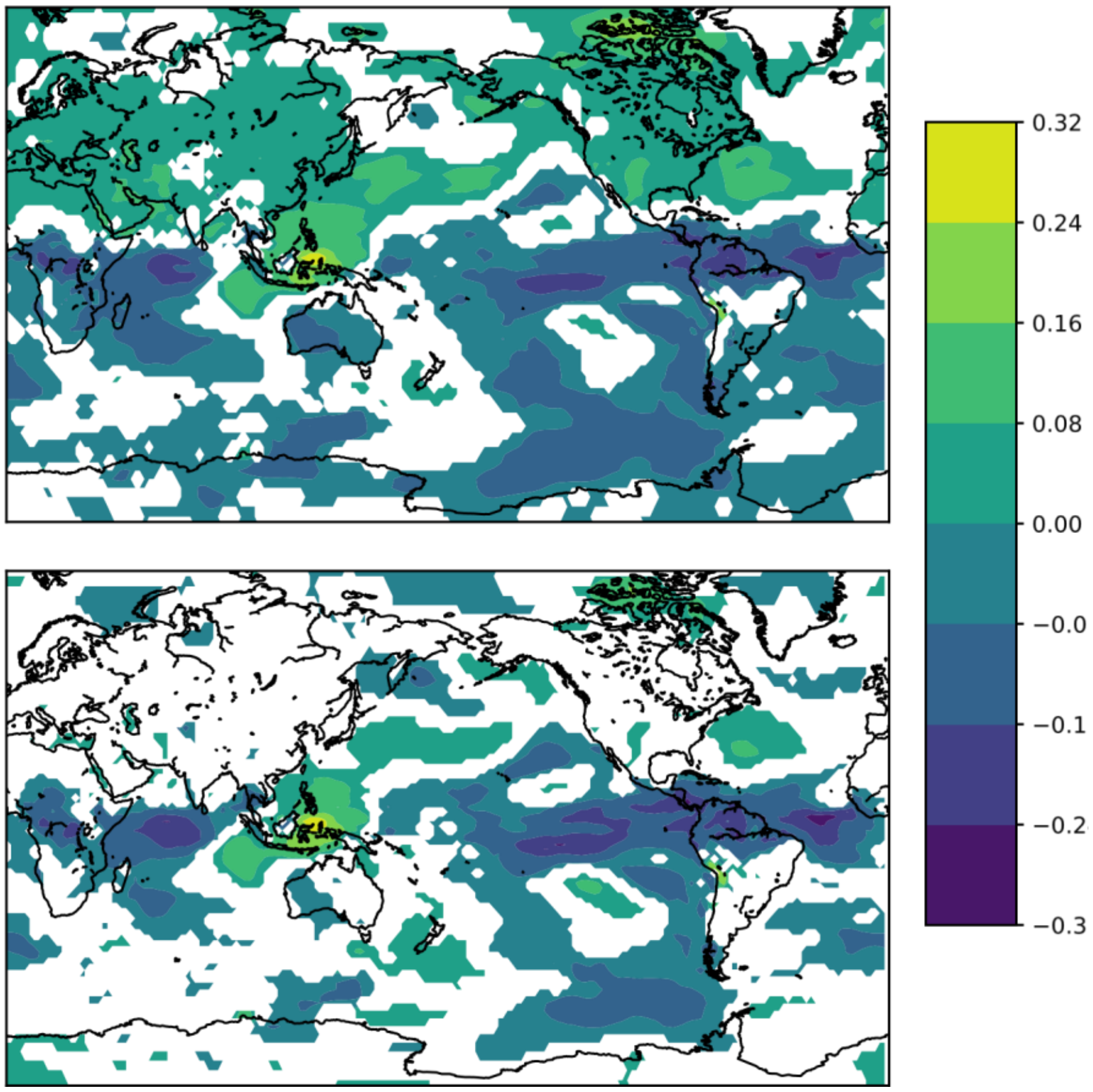

Figure 1: Spatial variation in significant correlation at 95% confidence interval between air temperature and (a) $R_z$ and (b) F10.7.

## 3. RESULTS AND DISCUSSION

The spatial variation of significant correlation of air temperature with $R_z$ and $F10.7$ during the entire solar cycle 23 is shown in Figure 1. The correlation of air temperature with $R_z$ showed hemispheric delineation. The northern hemisphere showed negative correlation with air temperature on both land and sea while a positive correlation was prevalent in the southern hemisphere. However, the hemispheric delineation was not obvious in the correlation of air temperature with $F10.7$ index. The highest negative correlations with $R_z$ index were found in the tropical region. This implies that $R_z$ play a significant role in the climatology of the tropics. Regions with high significant positive correlations between air temperature and $R_z$ include areas around Celebes sea (Pacific Ocean) and Queen Elizabeth Islands in North America. The correlation between $F10.7$ and air temperature were not found to be significant over major continental land masses except tropical land masses. The tropical land and oceans showed similar significant negative correlation with $F10.7$ index as in the $R_z$ index. The only significant positive correlation between air temperature and $F10.7$ index were also around Celebes sea (Pacific Ocean) and Queen Elizabeth Islands in North America.

The correlation between global air temperature and $R_z$ were also considered at the three seasons (Figure 2). In the J-season, significant negative correlations were predominant in many regions. Significantly high negative correlations were found across the equator except the Indian Ocean. Regions north and southeast of Australia showed significantly high positive correlations between air temper-ature and $R_z$ index. During this season, significant correlations were not found over the continental land masses except in Greenland and small regions in Africa and Europe. During the D-season, there were no significant negative correlations in the Pacific Islands, as in the J-season. However, significant high positive correlations were found in the Queen Elizabeth Islands, west of Greenland. Unlike the J-season, significant negative correlations were found in south Atlantic Ocean and Indian Ocean. Also, larger portion of North America showed significant correlations. The correlations around the Equator were found to be weaker and not predominant as in the J-season. Correlations between air temperature and $R_z$ during the E-Season were generally subdued with less spatial coverage compared to the J-season and D-season. The negative Equatorial correlations and positive Pacific Island correlations were observed to be weakest during this season. The correlation between air temperature and $F10.7$ index (Figure 3) at the three seasons showed identical patterns with the $R_z$ correlations.

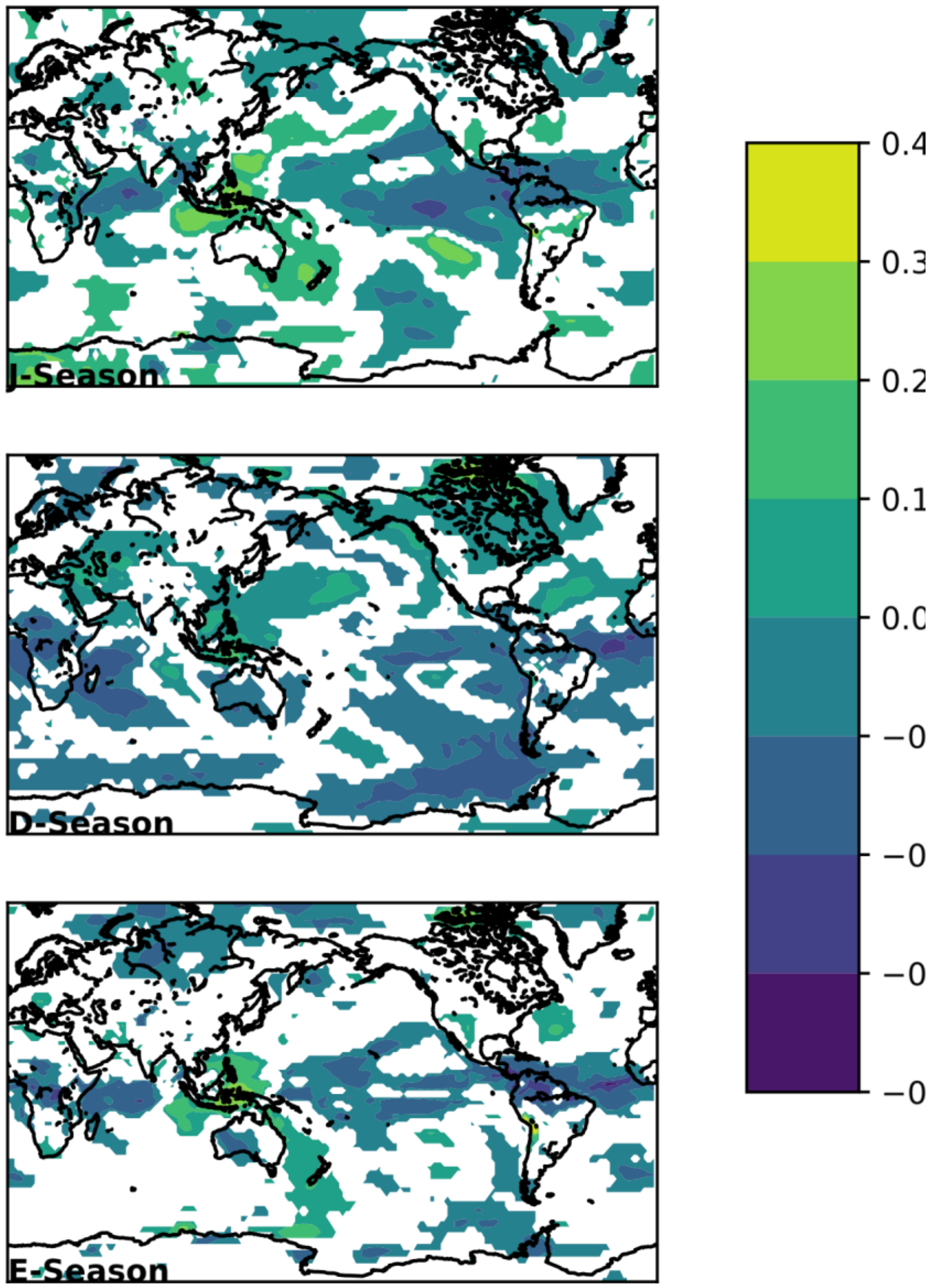

Figure 2: Seasonal spatial variation in significant correlation at 95% confidence interval between air temperature and $R_z$

In Figure 4, the annual correlation between global air temperature and $R_z$ were considered from 1997 to 2008. The correlation values were found to be in the range ±0.32. Hemispheric delineation were pronounced at the onset of the solar cycle from 1997 to 1999. Specifically, the northern hemisphere were observed to have predominantly significant positive correlations while the southern hemisphere we found to have prevalent significant negative correlations. In 2000, although the hemispheric delineation was present, they were observed only on continental land mass and the Arctic Ocean. However, in 2001 the correlations were only observed on large water bodies. From 2002 to 2005, there were sparse spatial distribution of significant correlations between air temperature and $R_z$ index. The year 2002 witnessed the sparsest spatial distribution of correlation values as only Equatorial Atlantic Ocean and a few other locations were found to be correlated. In 2003, the continental land mass of Africa, Europe, Australia, Asia, as well as North America and the Arctic sea showed significant correlation values. However, significant correlation values were not observed in the Arctic Ocean but over South America in 2004. In 2007 and 2008, there was an inversion of the hemispheric delineation. During 2007 and 2008, significant negative correlations were observed in the Northern Hemisphere while significant positive correlations were found in the Southern Hemisphere. However, the spatial distribution of the inverse hemispheric delineation was smaller in 2008 compared to 2007.

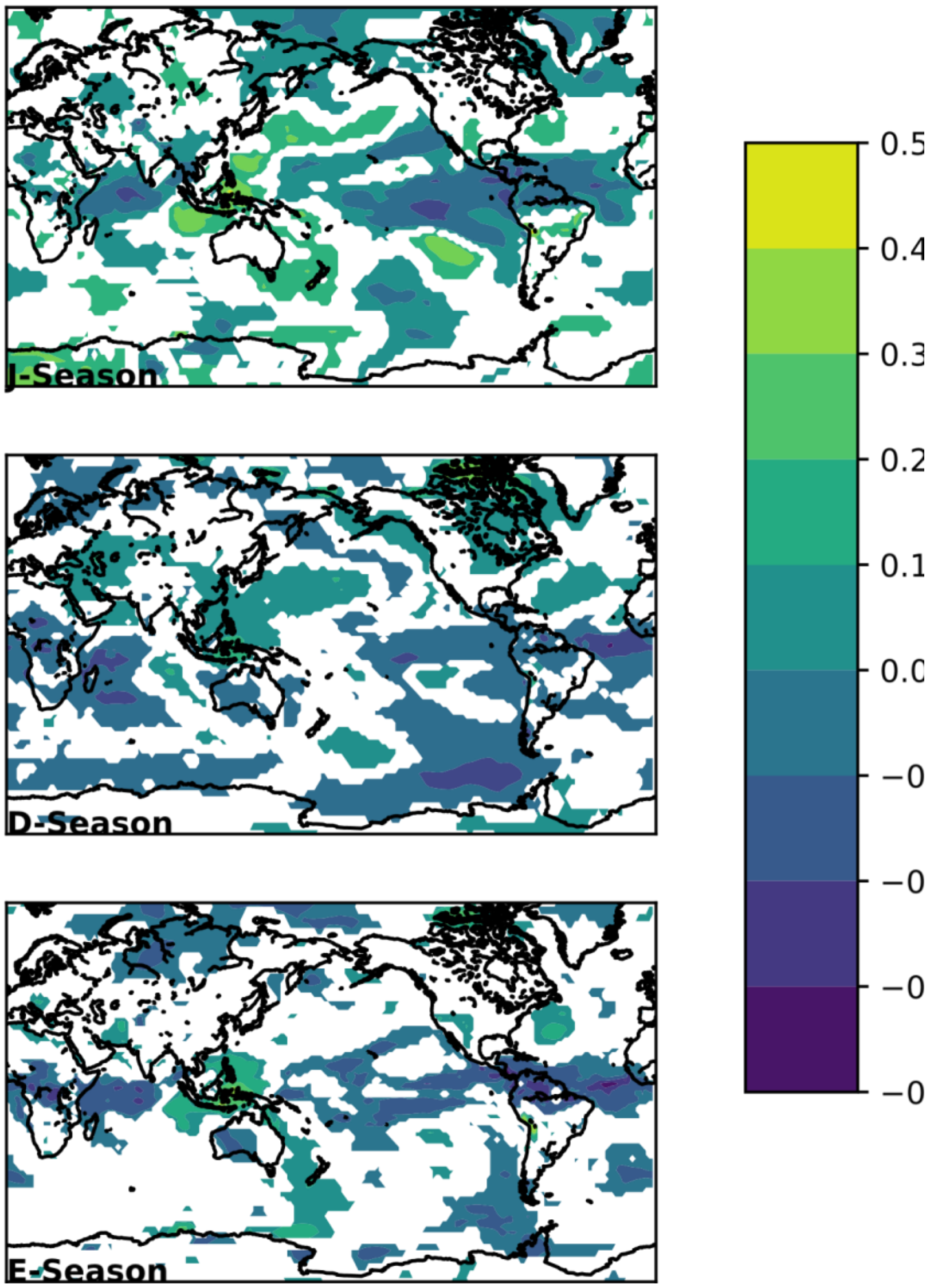

Figure 3: Seasonal spatial variation in significant correlation at 95% confidence interval between air temperature and $F10.7$

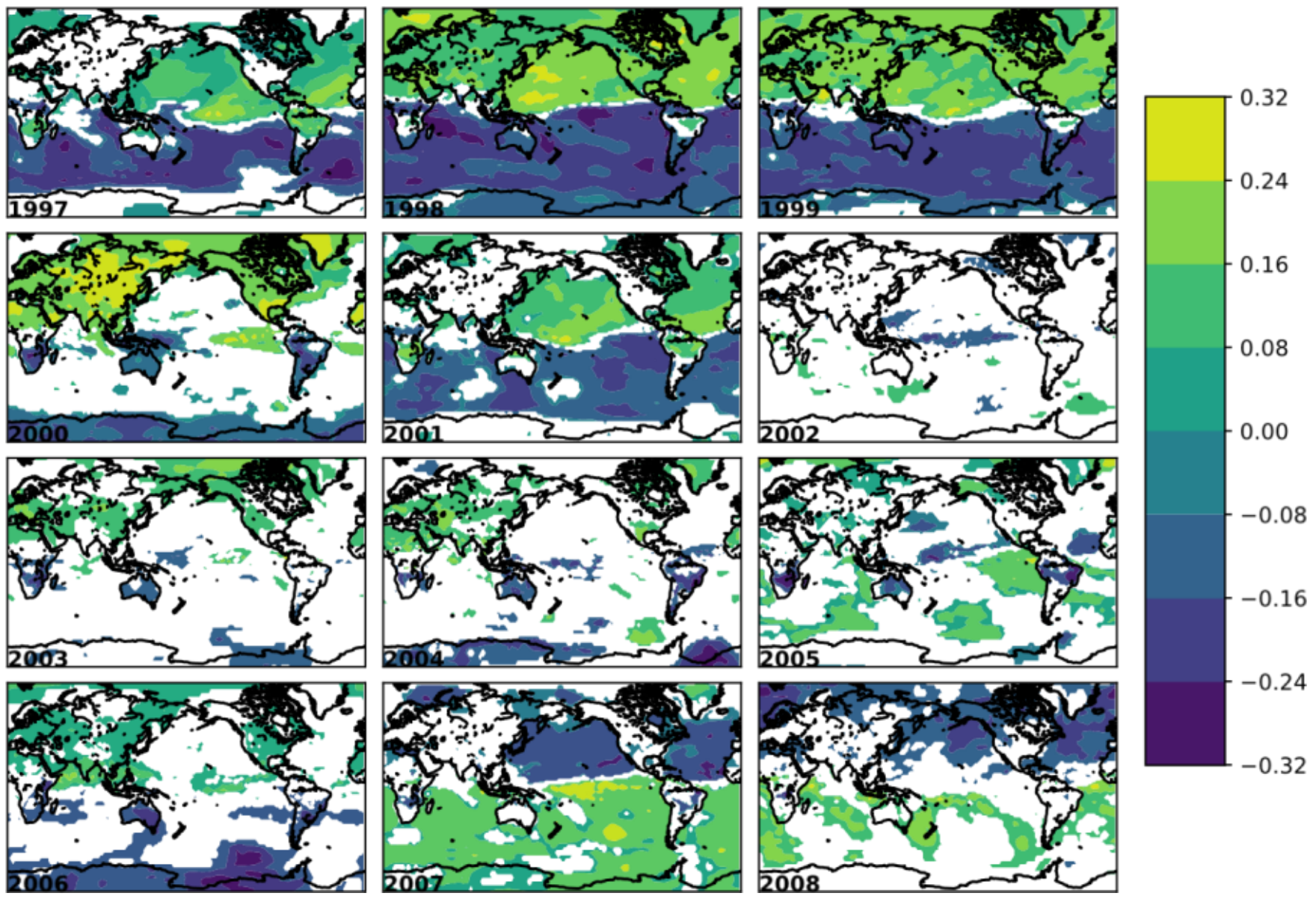

Figure 4: Spatial variation of significant correlation between $R_z$ and air temperature at 95% confidence interval for each year in solar cycle 23

Figure 5 showed the spatial correlation between global air temperature and $F10.7$ index for each year in solar cycle 23. The years 1997 to 1999 showed similar patterns as observed in the correlation with $R_z$ but with higher values. In 2001, the continental land mass in the northern hemisphere which did not show significant correlations under $R_z$ were found to exhibit negative correlations with $F10.7$ while the continental land mass in the southern hemisphere showed positive correlations. During the year 2002, an inverse hemispheric delineation was observed with negative correlations in the northern hemisphere and positive correlations in the southern hemisphere. This implies that $F10.7$ contribute more to global air temperature compared to $R_z$ index. Sparse spatial distribution of correlation values were also observed in the years 2003 to 2006, with 2005 showing the highest distribution. The correlation values in 2007 and 2008 also showed inverse hemispheric delineation but with larger spatial distribution compared with $R_z$.

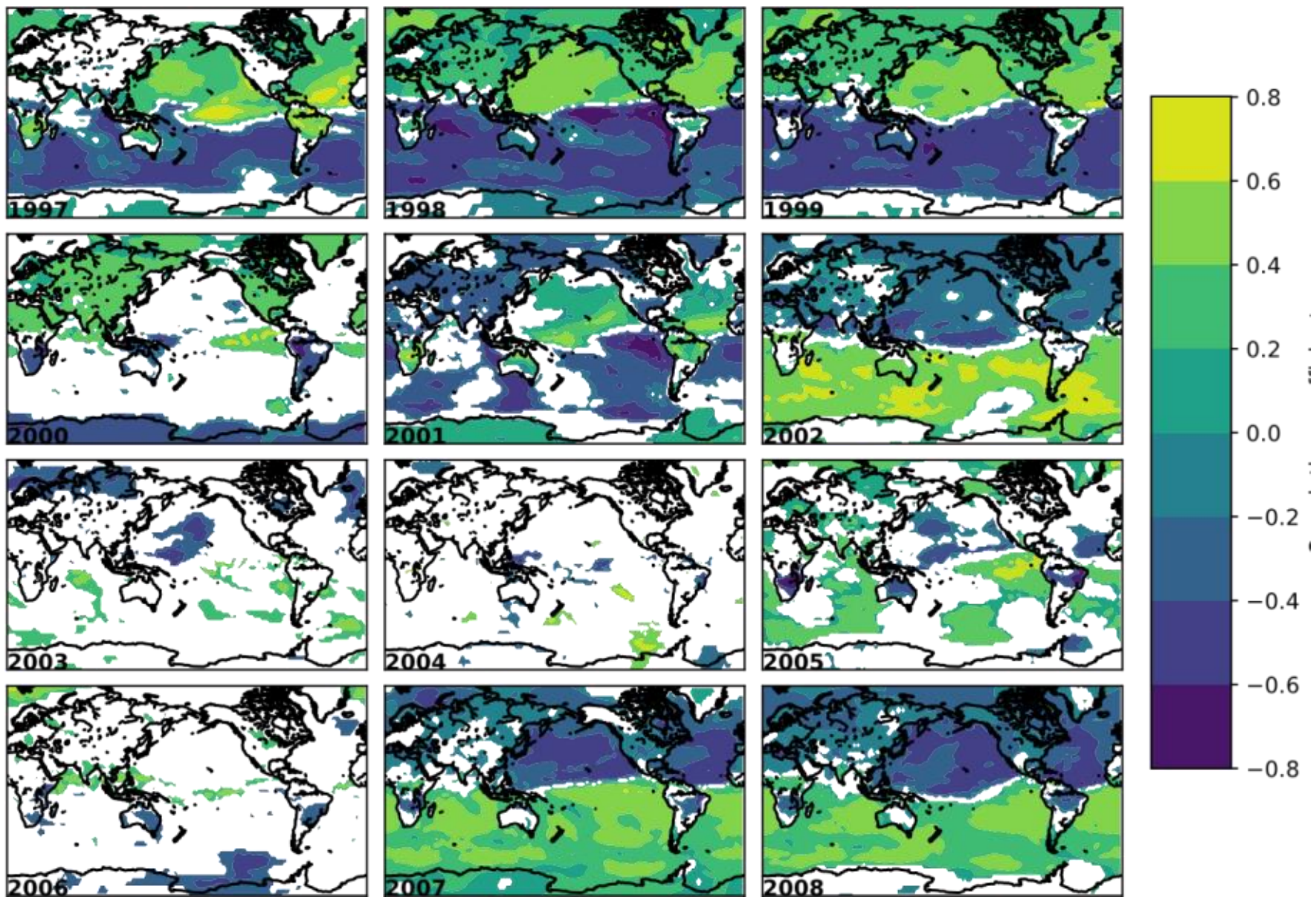

Figure 5: Spatial variation of significant correlation between $F10.7$ and air temperature at 95% confidence interval for each year in solar cycle 23

## 4. CONCLUSION

There has been no scientific consensus on the impact of solar activity on atmospheric weather. In this study, we have investigated the correlation between surface air temperature and two solar activity indices ($R_z$ and $F10.7$) on the global scale during solar cycle 23. This approach will help identify large scale patterns which can give more insight into the relationship between atmospheric weather and solar activity. Our study was conducted at seasonal, annual, and long term time scales for a clearer understanding. Our results showed hemispheric delineation at seasonal, annual, and long term considerations. Furthermore, while the years preceding the solar minimum showed preference for positive correlations in the north and negative correlations in the south, the receding years favours the opposite.

*Disclosures*

The authors declare no financial interests or conflict of interests in this manuscript.

*Data, Materials, and Code Availability*

Data used in this study is publicly available and links have been provided in the manuscript.